\documentclass{cjaa}

\usepackage{graphicx}                   
\input{epsf.sty}                        
\input{psfig.sty}                       

\begin{document}

  \title{The Dynamical Architecture and Habitable zones of the
Quintuplet Planetary System 55 Cancri
}

   \volnopage{Vol.0 (2008) No.2, 000--000}      
   \setcounter{page}{1}          

   \author{Jianghui JI
      \inst{1,2}\mailto{}
   \and Hiroshi Kinoshita
   \inst{3}
   \and Lin LIU
      \inst{4}
    \and Guangyu LI
     \inst{1,2}
    }
   \offprints{Jianghui Ji}                   

   \institute{Purple  Mountain  Observatory, Chinese  Academy
              of  Sciences, Nanjing  210008, China \\
             \email{jijh@pmo.ac.cn}
       \and  National Astronomical Observatories, Chinese Academy of Sciences,
             Beijing 100012, China \\
        \and
            National Astronomical Observatory,
            Mitaka, Tokyo 181-8588, Japan \\
        \and
            Department  of Astronomy, Nanjing University, Nanjing  210093, China\\
          }
   \date{Received~~2008/09/08}

\abstract{We perform  numerical simulations to study the secular
orbital evolution and dynamical structure in the quintuplet
planetary system 55 Cancri with the self-consistent orbital
solutions by Fischer and coworkers (2008). In the simulations, we
show that this system can be stable at least for $10^{8}$ yr.  In
addition, we extensively investigate the planetary configuration of
four outer companions with one  terrestrial planet in the wide
region of 0.790 AU $\leq a \leq $ 5.900 AU to examine the existence
of potential asteroid structure and Habitable Zones (HZs).  We show
that there are unstable regions for the orbits about 4:1, 3:1 and
5:2 mean motion resonances (MMRs) with the outermost planet in the
system, and several stable orbits can remain at 3:2 and 1:1 MMRs,
which is resemblance to the asteroidal belt in solar system. In a
dynamical point, the proper candidate HZs for the existence of more
potential terrestrial planets reside in the wide area between 1.0 AU
and 2.3 AU for relatively low eccentricities.
    }

   \authorrunning{Jianghui JI  et al.}
   \titlerunning{The Dynamical Structure in the 55 Cancri system}

   \maketitle

\section{Introduction}
The nearby star 55 Cancri is of spectral type K0/G8V with a mass of
$0.92 \pm 0.05 M_{\odot}$ (Valenti \& Fischer 2005).  Marcy et al.
(2002) reported a second giant planet with a long period of $\sim
14$ yr after the first planet discovered in 1997. The 55 Cnc system
can be very attractive, because first it hosts a distant giant
Jupiter-like planet about 5.5 AU resembling Jupiter in our solar
system. The second interesting thing is that this system may be the
only known planetary system in which two giant planets are close to
the 3:1 orbital resonance, and the researchers have extensively
studied the dynamics and formation of the 3:1 MMR in this system
(see Beaug{\'e} et al. 2003; Ji et al. 2003; Zhou et al. 2004; Kley,
Peitz, \& Bryden 2004; Voyatzis \& Hadjidemetriou 2006; Voyatzis
2008). Still, the additional collection of follow-up observations
and the increasing of precision of measurements (at present $\sim 1$
ms$^{-1}$ to $3$ ms$^{-1}$) have indeed identified more planets.
McArthur et al. (2004) reported the fourth planet with a small
minimum mass $\sim 14$ $M_{\oplus}$ that orbits the host star with a
short period of 2.8 day, by analyzing three sets of radial
velocities. The improvement of the observations will actually induce
additional discovery. Hence, it is not difficult to understand that
more multiple planetary systems or additional planets in the
multiple systems are to be dug out supplemental data.

More recently, Fischer et al. (2008) (hereafter Paper I) reported
the fifth planet in the 55 Cnc system with the Doppler shift
observations over 18 yrs, and showed that all five planets are in
nearly circular orbits and four have eccentricities under 0.10. It
is really one of the most extreme goals for the astronomers devoted
to searching for the extrasolar planets to discover a true solar
system analog, which may hold one or two gas giants orbiting beyond
4 AU that can be compared to Jupiter and Saturn in our own solar
system (Butler 2007, private communication; see also Gaudi et al.
2008). This indicates that several terrestrial planets may move in
the so-called Habitable Zones (HZs) (Kasting et al. 1993; Jones et
al. 2005), and the potential asteroidal structure can exist.
Considering the probability of the coplanarity and nearly circular
orbits for five planets (Paper I), the 55 Cnc system is suggested to
be a comparable twin of the solar system. Hence, firstly, in a
dynamical viewpoint, one may be concerned about the stability of the
system over secular timescale. On the other hand, the small bodies
as terrestrial objects may exist in this system and are to be
detected with forthcoming space-based missions (\textit{Kepler,
SIM}).  In this paper,  we focus on understanding the dynamical
structure and finding out suitable HZs for life-bearing terrestrial
planets in this system.

\section{Dynamical Analysis}
In this paper, we adopt the orbital parameters of the 55 Cancri
system provided by Paper I (see their Table 4). For the convenience
of narration, we re-label the planets according to the ascendant
semi-major axes in the order from the innermost to the outermost
planet (e.g., B, C, D, E, F), while the original names discovered in
the chronological order are also accompanied but in braces (see
Table 1). Furthermore, McArthur et al. (2004) derived the orbital
inclination $i = 53^{\circ} \pm 6^{\circ}.8$ with respect to the
sight line for the outermost planet from \textit{HST} astrometric
data by measuring the apparent astrometric motion of the host star.
In the simulations, we adopt this estimated orbital inclination of
$53^{\circ}$ and further assume all the orbits to be coplanar. With
the planetary masses $M \sin i$ reported in Table 1, then we obtain
their true masses. Specifically, the masses of five planets are
respectively, 0.03 $M_{Jup}$, 1.05 $M_{Jup}$, 0.21 $M_{Jup}$, 0.18
$M_{Jup}$ and 4.91 $M_{Jup}$, where $\sin i = \sin 53^{\circ} =
0.7986$. Thus, we take the stellar mass $M_{c}$ of 0.94 $M_{\odot}$
(Paper I), and the planetary masses above-mentioned in the numerical
study, except where noted. We utilize N-body codes (Ji, Li \& Liu
2002) to perform numerical simulations by using RKF7(8) and
symplectic integrators (Wisdom \& Holman 1991) for this system. In
the numerical runs, the adopted time stepsize is usually $\sim$ 2\%
- 5\% of the orbital period of the innermost planet. In addition,
the numerical errors were effectively controlled over the
integration timescale, and the total energy is generally conserved
to $10^{-6}$ for the integrations. The typical timescale of
simulations of the 55 Cnc system is from 100 Myr to 1 Gyr.

\subsection{The Stability of the 55 Cancri Planetary System}
\subsubsection{case 1: 5-p for $10^8$ yr}
To explore the secular stability of this system, firstly, we
numerically integrated the five-planet system on a timescale of
$10^{8}$ yr, using the initials listed in Table 1. In Figure 1, a
snapshot of the secular orbital evolution of all planets is
illustrated, where $Q_{i}=a_{i}(1 + e_{i})$, $q_{i}=a_{i}(1 -
e_{i})$ (the subscript $i=1-5$, individually, denoting Planet B, C,
D, E, and F) are, respectively, the apoapsis and periapsis
distances. In the secular dynamics, the semi-major axis $a_{1}$ and
$a_{2}$ remain unchanged to be 0.0386 and 0.115 AU, respectively,
for $10^{8}$ yr, while $a_{3}$, $a_{4}$ and $a_{5}$ slightly librate
about 0.241, 0.786, and 6.0 AU with quite small amplitudes over the
same timescale. The variations of eccentricities during long-term
evolution are followed, where $0.23 < e_{1}< 0.28$, $0.0 < e_{2}<
0.03$, $0.034 < e_{3}< 0.069$, $0.0 < e_{4}< 0.013$, and $0.056 <
e_{5}< 0.095$, implying that all the eccentricities undergo
quasi-periodic modulations. In Fig.1, we note that the time
behaviors of $Q_{i}$ and $q_{i}$ show regular motions of bounded
orbits for all five planets and indicate their orbits are well
separated during the secular evolution due to small mutual
interactions, which again reflect the regular dynamics of the
eccentricities over secular timescale. In the numerical study, we
find the system can be dynamically stable and last at least for
$10^{8}$ yr. Thus our numerical outcomes strengthen and verify those
of Paper I for the integration of $10^6$ yr, which also showed the
system can remain stable over 1 Myr and the variations of all
planetary eccentricities are modest.

Secondly, we further performed an extended integration for the
planetary configuration simply consisting of four outer planets over
timescale up to 1 Gyr (see Figure 2). The longer integration again
reveals that the orbital evolution of four planets are quite similar
to those exhibited in Fig.1, and then strongly supports the secular
stability of this system. In a recent study, Gayon et al. (2008)
show that the 55 Cnc system may remain a stable chaos state as the
planetary eccentricities do not grow over longer timescale.
Therefore, it is safely to conclude that the 55 Cnc system remain
dynamically stable in the lifetime of the star.

In order to assess the stability of 55 Cnc with respect to the
variations of the planetary masses, we first fix $\sin i$ in
increment of 0.1 from 0.3 to 0.9. In the additional numerical
experiments, we simply vary the masses but keep all orbital
parameters (Table 1), again restart new runs of integration for the
five-planet system for 100 - 1000 Myr with the rescaled masses. As a
result, we find the system could remain definitely stable for the
above investigated timescale with slight vibrations in semi-major
axes and eccentricities for all planets, indicating the present
configuration is not so sensitive to the planetary masses.
Subsequently, we again examine the stabilities of different orbital
configurations within the error range of the Keplerian orbital fit
given by Paper I. Herein 100 simulations are carried out for 10 Myr,
and the numerical results show that all the runs are stable over the
simulation timescale, indicating that this five-planet system is
fairly robust with respect to the variational planetary
configurations.

\subsubsection{case 2: 7-p for $10^8$ yr}
However, Paper I argued that the 6th or more planets could exist and
maintain dynamical stability in the large gap between Planets E and
F in this system. Next, we also integrate the 55 Cnc system with
additional planets (2 massive terrestrial planets, Earth at 1 AU and
Mars at 1.52 AU) to mimic the situation of the inner solar system.
In this runs, we examine the configuration consisting of 5 planets
and 2 terrestrial bodies to study the coexistence of multiple
objects. This means that we directly place Earth and Mars into the
55 Cnc system to simulate "the inner solar system", where the
orbital elements for above terrestrial planets are calculated from
JPL planetary ephemerides DE405 at Epoch JD 2446862.3081
corresponding to the outermost companion (see Table 2), e.g., the
semi-major axes are respectively, 1.00 and 1.524 AU. The five
planets are always assumed to be coplanar in the simulations, thus
the inclinations for 2 terrestrial planets refer to the fundamental
plane of their orbits. In this numerical experiment, we find that
the 7-p system can remain dynamically stable and last at least for
$10^{8}$ yr. In Figure 3 are shown the time behaviors of $Q$
(\textit{yellow line}) and $q$ (\textit{black line}) for Mars, Earth
and Planet E. The numerical results show the regular bounded motions
that their semi-major axes and eccentricities do not dramatically
change in their secular orbital evolution, and this is also true for
the other four planets in the 55 Cnc. It is not so surprised for one
to realize that two additional terrestrial planets could exist for
long time because the gravitational perturbations arise from other
planets are much smaller. In the following section, we will
extensively explore this issue on the dynamical architecture for the
Earth-like planets in the system.

\section{Dynamical Architecture and Potential HZs}
To investigate the dynamical structure and potential HZs in this
system, we extensively performed additional simulations with the
planetary configuration of coplanar orbits of four outer companions
with one  terrestrial planet. In this series of runs, the mass of
the assumed terrestrial planet selected randomly in the range 0.1
$M_{\oplus}$ to 1.0 $M_{\oplus}$. The initial orbital parameters are
as follows: the numerical investigations were carried out in [$a,
e$] parameter space by direct integrations, and for a uniform grid
of 0.01 AU in semi-major axis (0.790 AU $\leq  a \leq $ 5.900 AU)
and 0.01 in eccentricity ($0.0 \leq e \leq 0.2$), the inclinations
are $0^{\circ} < I < 5^{\circ}$. The angles of the nodal longitude,
the argument of periastron, and the mean anomaly are randomly
distributed between $0^{\circ}$ and $360^{\circ}$ for each orbit.
Then each terrestrial mass body was numerically integrated with four
outer planets in the 55 Cnc system. In total, about 10,750
simulations were exhaustively run for typical integration time spans
from $10^{5}$ to $10^{6}$ yr (about $10^{6}$ - $10^{7}$ times the
orbital period of Planet C). Then, our main results now follow.

Figure 4 shows the contours of the surviving time for Earth-like
planets (\textit{Upper}) and the status of their final
eccentricities (\textit{Lower}) for the integration over $10^{5}$
yr, and the horizontal and vertical axes represent initial $a$ and
$e$ of the orbits. Fig. 4 (\textit{Upper}) displays that there are
stable zones for the Earth-like planets in the region between 1.0
and 2.3 AU with final low eccentricities of $e < 0.10$. The extended
simulations ($10^{6}$ yr) for the objects of the above region also
exhibit the same results. This zone may be strongly recommended to
be one of the potential candidate HZs in the 55 Cnc, and our results
coincide with those by Jones et al. (2005), who showed the possible
HZs of 1.04 AU $< a < $ 2.07 AU. Still, the outcomes presented here
have confirmed those in \S2.1.2, where we show that the stable
configuration of Earth at 1.00 AU and Mars at 1.523 AU in this
five-planet system. The sixth planet or additional habitable bodies
may be expected to revealed in this region by future
observations\footnote{The semi-amplitude of wobble velocity $K
\propto {M_{p} \sin i}/ {\sqrt{a(1-e^2)}}$ (with $M_{p}\ll M_{c}$),
herein $M_{c}$, $M_{p}$, $a$, $e$ and $i$ are, respectively, the
stellar mass, the planetary mass, the orbital semi-major axis, the
eccentricity and the inclination of the orbit relative to the sky
plane. This means that planets with larger masses and (or) smaller
orbits could have larger $K$. For example, a planet of 1.0
$M_{\oplus}$ at 1 AU in a nearly circular orbit may cause stellar
wobble about 0.10 m/s. In this sense, much higher Doppler precision
is required to discover such Earth-like planets in future.}.

In general, the planetary embryos or planetesimals may be possibly
captured into the mean motion resonance regions or thrown into HZs
by a giant planet under migration due to the planet-disk interaction
and could survive during the final planetary evolution over the
secular timescale after complex scenarios of secular resonance
sweeping, gravitationally scattering, and late heavy bombardments
(Nagasawa et al. 2005; Thommes et al. 2008). We note that there are
strongly unstable orbits \footnote{We define an unstable orbit as an
Earth-like planet is ejected far away or moves too close to the
parent star or the giant planets, meeting the following criteria:
(1) the eccentricity approaches unity, (2) the semi-major axis
exceeds a maximum value, e.g., 1000 AU, (3) the assuming planet
collides with the star or enters the mutual Hill sphere of the known
giant planets.} for the low-mass planets initially distributed in
the region $3.9$ AU $< a < 5.9$ AU, where the planetary embryos have
very short dynamical surviving time. In the meantime, the
eccentricities can be quickly pumped up to a high value $\sim 0.9$
(see Fig. 4, \textit{Lower}). We note that the orbital evolution is
not so sensitive to the initial masses. In fact, these planetary
embryos are involved in many of MMRs with the outermost giant in the
55 Cnc system, e.g., 7:4 (4.063 AU) and 3:2 MMRs (4.503 AU). The
overlapping resonance mechanism (Murray \& Dermott 1999) can reveal
their chaotic behaviors of being ejected from the system in short
dynamical lifetime $\sim 10^2 - 10^3$ yr, furthermore the majority
of orbits are within the sphere of  3 times Hill radius
($R_{H}={\left({M_{5}}/(3M_{c})\right)}^{1/3}a_{5}$,
$3R_{H}\doteq2.10$ AU) of the 14-yr planet. Using resonance
overlapping criterion (MD99; Duncan et al. 1989), the separation in
semi-major axis
$\Delta{a}\approx1.5{\left({M_{5}}/M_{c}\right)}^{2/7}a_{5}\doteq1.95$
AU, then the inner boundary $R_{O}= a_{5} - \Delta{a} $ for Planet F
is at $\sim 3.95$ AU. And the orbits in this zone become chaotic
during the evolution because the planets are both within 3 $R_{H}$
and in the vicinity of $R_{O}$. Similarly, there exist unstable
zones for the nearby orbits around Planet E (0.78 AU $< a <$ 0.90
AU), which may not be habitable in dynamical point.

It is suggested that MMRs can play an important role in determining
the orbital dynamics of the terrestrial bodies, which are either
stabilized or destabilized in the vicinities of the MMRs. The
outermost giant, like Jupiter, may shape and create the
characteristic of dynamical structure of the small bodies. Most of
the initial orbits for planetary embryos located about 3:1 (2.837
AU), 5:2 (3.204 AU), and 4:1 MMRs (2.342 AU), are quickly cleared
off by the perturbing from Planet F.  In the region of $2.4$ AU $< a
< 3.8$ AU, stable zones are separated by the  mean motion resonance
barriers, e.g., 3:1 and 5:2 MMRs. Note that the initial orbits for
the relatively low eccentricity (under 0.06) for 4:1 MMR can remain
stable over the simulation timescale. However, the terrestrial
bodies about 7:3 MMR (3.354 AU) and 2:1 MMR (3.717 AU) are both on
the edge of the stability, and the former are close to 5:2 MMR,
while the latter just travel around the inner border of $3R_{H}$ at
$\sim 3.80$ AU. The extended longer integrations show that their
eccentricities can be further excited to a high value and a large
fraction of them lose stabilities in the final evolution. The above
gaps are apparently resembling those of the asteroidal belt in solar
system. In the simulations, several stable orbits can be found about
3:2 MMR at 4.503 AU, which is analogous to the Hilda group for the
asteroids in the solar system, surviving at least for $10^{6}$ yr.
In addition, the other several stable cases are the so-called Trojan
planets (1:1 MMR), residing at $\sim $ 5.9 AU. The studies show that
the stable Trojan configurations may be possibly common in the
extrasolar planetary systems (Dvorak et al. 2004; Ji et al. 2005;
Gozdziewski \& Konacki 2006). Indeed, terrestrial Trojan planets
with circular orbits $\sim$ 1 AU could potentially be habitable, and
are worthy of further investigation in future.

\section{Summary and Discussions}
In this work, we have studied the secular stability and dynamical
structure and HZs of the 55 Cnc planetary system. We now summarize
the main results as follows:

(1) In the simulations, we show that the quintuplet planetary system
can remain dynamically stable at least for $10^{8}$ yr and that the
stability would not be greatly influenced by shifting the planetary
masses. Account for the nature of near-circular well-spaced orbits,
the 55 Cnc system may be a close analog of the solar system. In
addition, we extensively investigated the planetary configuration of
four outer companions with one terrestrial planet in the region
0.790 AU $\leq a \leq $ 5.900 AU to examine the existence of
potential Earth-like planets and further study the asteroid
structure and HZs in this system. We show that unstable zones are
about 4:1, 3:1 and 5:2 MMRs in the system, and several stable orbits
can remain at 3:2 and 1:1 MMRs.  The simulations not only present a
clear picture of a resembling of the asteroidal belt in solar
system, but also may possibly provide helpful information to
identify the objects when modeling multi-planet orbital solutions
(Paper I) by analyzing RV data. The dynamical examinations are
helpful to search for best-fit stable orbital solutions to consider
the actual role of the resonances, where some of best-fit solutions
close to unstable islands of MMRs can be dynamically ruled out in
the fitting process. As well-known, the extensive investigations in
the planetary systems (Menou \& Tabachnik 2003; {\'E}rdi et al.
2004; Ji et al. 2005, 2007; Pilat-Lohinger et al. 2008; Raymond et
al. 2008) show that the dynamical structure is correlated with mean
motion and secular resonances. The eccentricities of the
planetesimals (or terrestrial planets) can be excited by sweeping
secular resonance (Nagasawa \& Ida 2000) as well as mean motion
resonances, thus the orbits of the small bodies can undergo mutual
crossings and then they are directly cleared up in the
post-formation stage. In conclusion, the mentioned dynamical factors
perturbing from the giant planets will influence and determine the
characteristic distribution of the terrestrial planets in the late
stage formation of the planetary systems, to settle down the
remaining residents in the final system.

(2) As the stellar luminosity of 55 Cnc is lower than that of the
Sun, the HZ should shift inwards compared to our solar system. It
seems that the newly-discovered planet at $\sim 0.783$ AU can reside
in the HZs (Rivera \& Haghighipour 2007), and this planet may be
habitable provided that it bears surface atmosphere to sustain the
necessary liquid water and other suitable life-developing conditions
(Kasting et al. 1993). In a dynamical consideration, the proper
candidate HZs for the existence of more potential terrestrial
planets reside in the wide area between 1.0 AU and 2.3 AU for
relatively low eccentricities, and the maintenance of low
eccentricity can play a vital role in avoiding large seasonal
climate variations (Menou \& Tabachnik 2003) for the dynamical
habitability of the terrestrial planets. Moreover, our numerical
simulations also suggest that additional Earth-like planets
(\S2.1.2) can also coexist with other five known planets in this
system over secular timescale. This should be carefully examined by
abundant measurements and space missions (e.g. \textit{Kepler} and
\textit{TPF}) for this system in future.

\begin{acknowledgements}
We would like to thank the anonymous referee for valuable comments
and suggestions that help to improve the contents. We are grateful
to G.W. Marcy  and D. A. Fischer for sending us their manuscript and
insightful discussions. This work is financially supported by the
National Natural Science Foundations of China (Grants 10573040,
10673006, 10833001, 10203005) and the Foundation of Minor Planets of
Purple Mountain Observatory. We are also thankful to Q.L. Zhou for
the assistance of computer utilization. Part of the computations
were carried out on high performance workstations at Laboratory of
Astronomical Data Analysis and Computational Physics of Nanjing
University.
\end{acknowledgements}

\clearpage

\begin{table}[]
  \caption[]{The orbital parameters of 55 Cancri planetary system\inst{a}.}
  \label{Tab1}
  \begin{center}
  \begin{tabular}{lclcclc}
  \hline\noalign{\smallskip}
Planet &$M$sin$i$($M_{Jup}$) &$P$(days)  &$a$(AU) &$e$ &$\varpi$(deg) &$T_{p}$ \\
  \hline\noalign{\smallskip}

Planet B (e) & 0.0241  &2.796744  &0.038 & 0.2637    &156.500   & 2447578.2159 \\
Planet C (b) & 0.8358  &14.651262 &0.115 & 0.0159    &164.001   & 2447572.0307 \\
Planet D (c) & 0.1691  &44.378710 &0.241 & 0.0530    & 57.405   & 2447547.5250 \\
Planet E (f) & 0.1444  &260.6694  &0.785 & 0.0002    &205.566   & 2447488.0149 \\
Planet F (d) & 3.9231  &5371.8207 &5.901 & 0.0633    &162.658   &2446862.3081 \\

  \noalign{\smallskip}\hline
  \end{tabular}
  \end{center}
  {\textit{a}. The parameters are taken from Table 4 of Fischer
  et al. (2008). The mass of the star is 0.94 $M_{\odot}$.}
\end{table}

\begin{table}[]
  \caption[]{The orbital elements for 2
   terrestrial planets at JD 2446862.3081 (From DE405).}
  \label{Tab2}
  \begin{center}
  \begin{tabular}{lccllll}
  \hline\noalign{\smallskip}
Planet &$a$ (AU)&$e$ &$I$(deg) &$\Omega$(deg)& $\omega$(deg) &M(deg) \\
  \hline\noalign{\smallskip}

  Earth  &1.000   &0.0164   &0.002  &348.33  &115.231  & 61.647  \\
  Mars   &1.524   &0.0935   &1.850  &49.60   &286.352  & 85.614 \\

  \noalign{\smallskip}\hline
  \end{tabular}
  \end{center}
\end{table}
\clearpage

\begin{figure}[htbp]

\centerline{\includegraphics[width=0.90\textwidth]{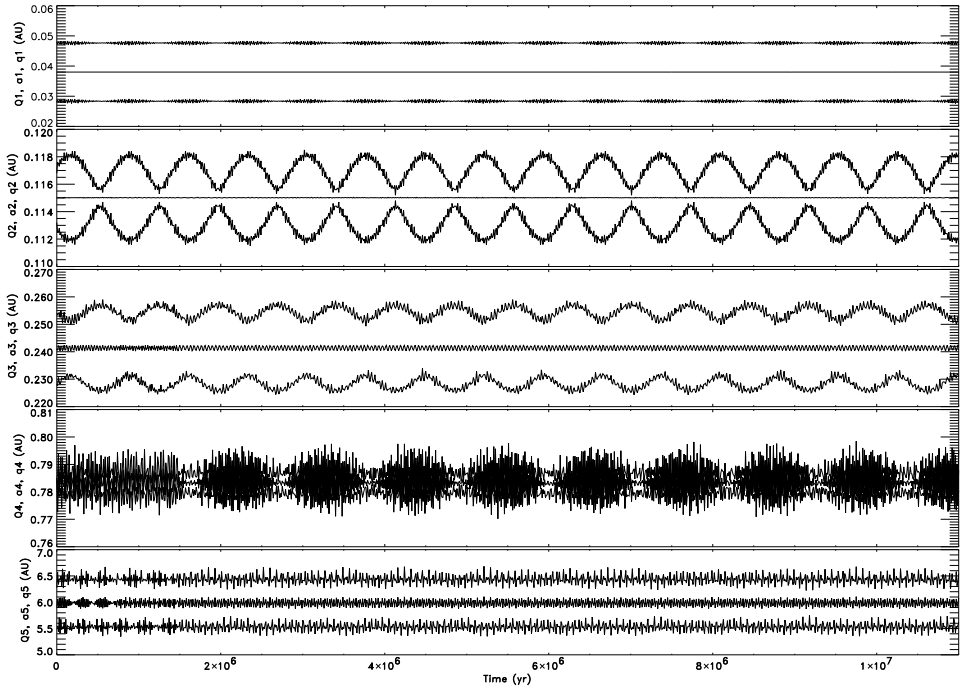}}
\caption{\normalsize Snapshot of the secular orbital evolution of
all planets is illustrated, where $Q_{i}=a_{i}(1 + e_{i})$,
$q_{i}=a_{i}(1 - e_{i})$ (the subscript $i=1-5$, each for Planet B,
C, D, E, and F) are, respectively, the apoapsis and periapsis
distances. $a_{1}$ and $a_{2}$ remain unchanged to be 0.0386 and
0.115 AU, respectively, while $a_{3}$, $a_{4}$ and $a_{5}$ slightly
librate about 0.241, 0.786, and 6.0 AU with smaller amplitudes for
$10^{8}$ yr. The simulations indicate the secular stability of the
55 Cnc.} \label{fig1}
\end{figure}

\begin{figure}[htbp]
\centerline{\includegraphics[width=0.90\textwidth]{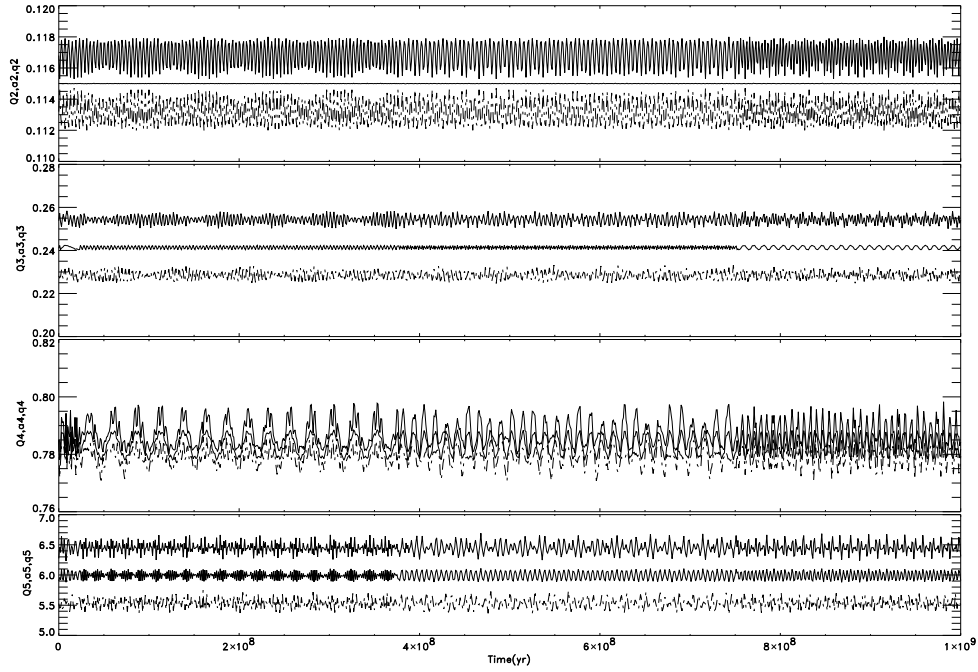}}
\caption{\normalsize  Numerical simulations for four outer planets
for $10^{9}$ yr. The apoapsis and periapsis distances:
$Q_{i}=a_{i}(1 + e_{i})$, $q_{i}=a_{i}(1 - e_{i})$ (the subscript
$i=2-5$, each for Planet C, D, E, and F). The long-term simulation
shows that the system can remain stable over 1 Gyr.} \label{fig2}
\end{figure}

\begin{figure}[htbp]
\centerline{\includegraphics[width=0.90\textwidth]{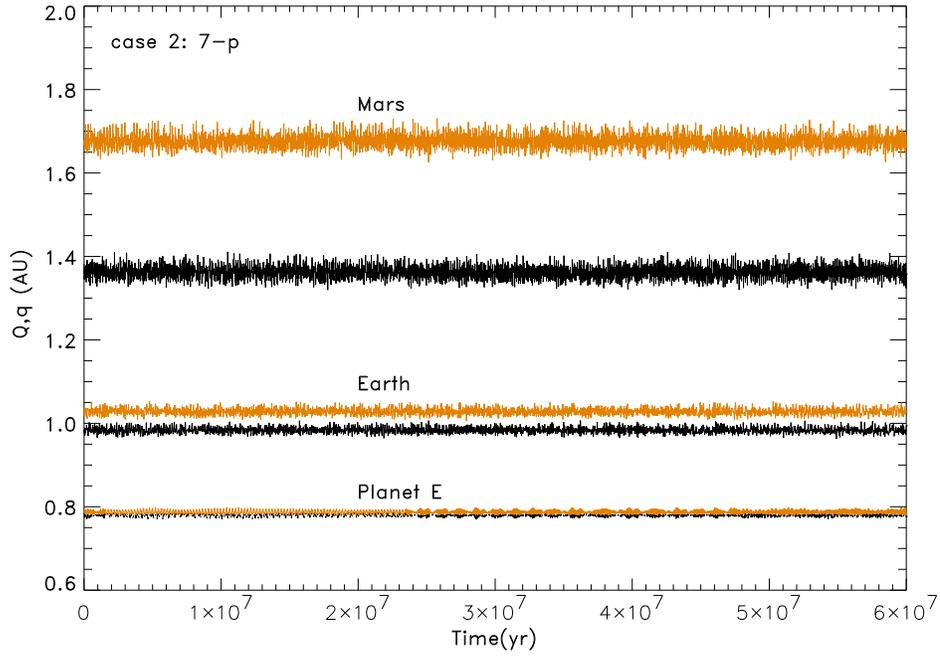}}
\caption{\normalsize Simulation for 7-p case. The system can be
dynamically stable and last at least for $10^{8}$ yr, the time
behaviors of $Q$ (\textit{yellow line}) and $q$ (\textit{black
line}) each for Mars, Earth and Planet E. The results show that
their semi-major axes and eccentricities do not dramatically change
in the secular orbital evolution, and it is also true for the other
four planets in the 55 Cnc.} \label{fig3}
\end{figure}

\begin{figure}[htbp]
\centerline{\includegraphics[width=0.90\textwidth]{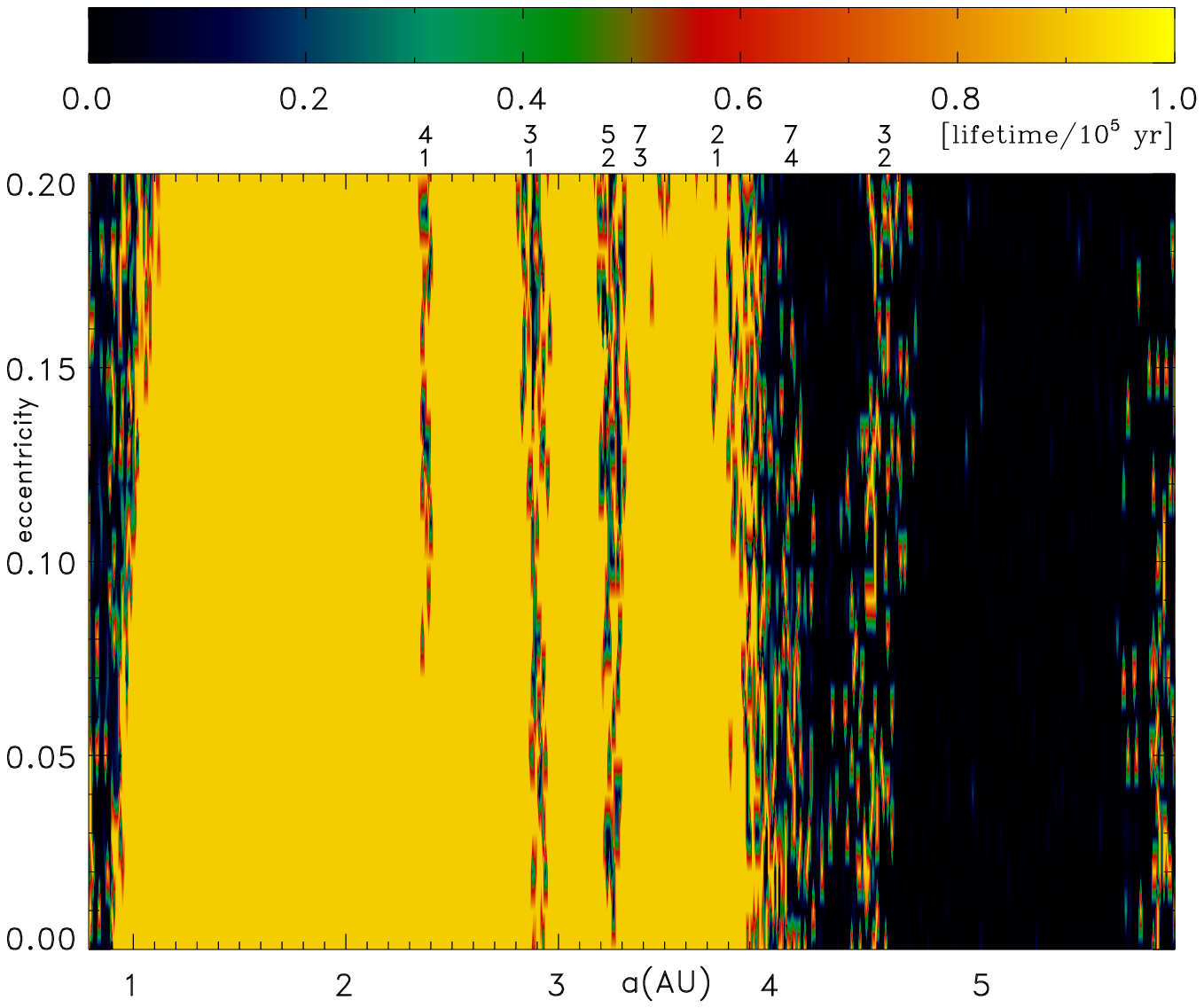}}
\centerline{\includegraphics[width=0.90\textwidth]{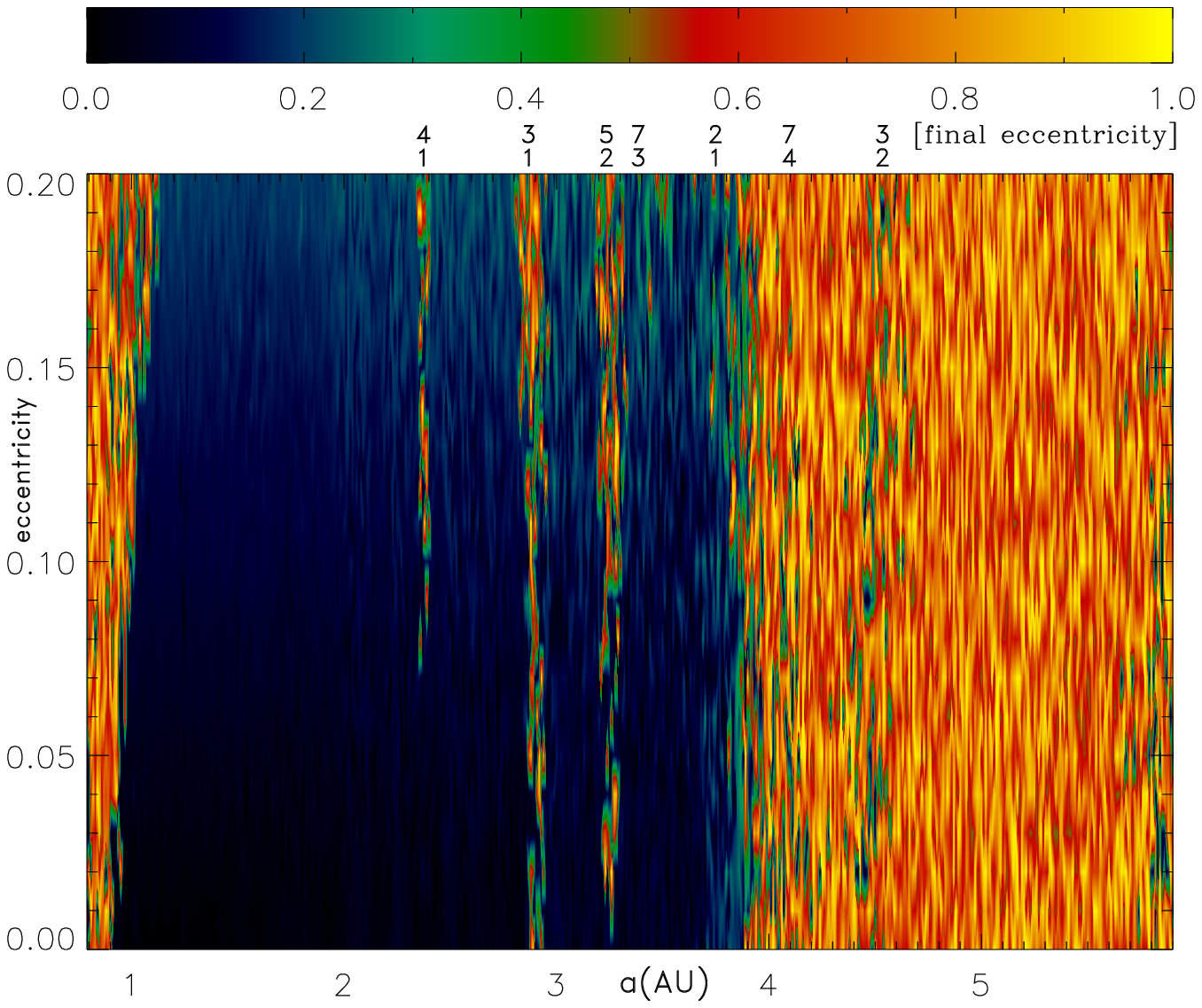}}

\caption{{\normalsize \textit{Upper}: Contour of the surviving time
for Earth-like planets for the integration of $10^{5}$ yr.
\textit{Lower}: Status of their final eccentricities. Horizontal and
vertical axes are the initial $a$ and $e$. Stable zones for the
Earth-like planets in the region between 1.0 and 2.3 AU with final
low eccentricities of $e < 0.10$. Unstable islands, e.g., 3:1 and
5:2 MMRs, have separated the region of $2.4$ AU $< a < 3.8$ AU.
Strongly chaos happen for the low-mass bodies initially distributed
in $3.9$ AU $< a < 5.9$ AU,  and their eccentricities can be quickly
pumped up to a high value $\sim 0.9$.}} \label{fig4}
\end{figure}

\label{lastpage}

\begin{thebibliography}{}

\bibitem[Beaug{\'e} et al.(2003)]{Bea03}
Beaug{\'e}, C., et al. \ 2003, \apj, 593, 1124


\bibitem[Dvorak et al. (2004)]{Dvo04}
Dvorak, R., et al. 2004, \aap, 426, L37

\bibitem[Duncan et al.(1989)]{Dun89}
Duncan, M., Quinn, T., \& Tremaine, S.\ 1989, Icarus, 82, 402

\bibitem[{\'E}rdi et al.(2004)]{Erd04}
{\'E}rdi, B., et al. \ 2004, \mnras, 351, 1043

\bibitem[Fischer et al.(2008)]{Fis08}
Fischer, D.~A., et al. 2008, \apj, 675, 790 (Paper I)

\bibitem[Gaudi et al.(2008)]{Gau08}
Gaudi, B.~S., et al.\ 2008, Science, 319, 927

\bibitem[Gayon et al.(2008)]{Gay08}
Gayon, J., Marzari, F., \& Scholl, H.\ 2008, \mnras, 389, L1

\bibitem[Gozdziewski(2006)]{Goz06}
Gozdziewski, K., \& Konacki, M.\ 2006, \apj, 647, 573

\bibitem[Ji et al.(2002)]{Ji02}
Ji, J.~H., Li, G.Y., \& Liu, L. 2002, \apj, 572, 1041

\bibitem[Ji et al.(2003)]{Ji03}
Ji, J.~H., Kinoshita, H.,  Liu, L., \&  Li, G.Y.  2003, \apj, 585,
L139

\bibitem[Ji et al.(2005)]{Ji05}
Ji, J.~H., Liu, L., Kinoshita, H., \& Li, G.Y.  2005, \apj, 631,
1191

\bibitem[Ji et al.(2007)]{Ji07}
Ji, J.~H., Kinoshita, H., Liu, L., \& Li, G.Y.  2007, \apj, 657,
1092

\bibitem[Jones et al.(2005)]{Jon05}
Jones, B.~W., et al. \ 2005, \apj, 622, 1091

\bibitem[Kasting et~al.(1993)]{Kas93}
Kasting, J. F., et al. 1993, Icarus, 101, 108

\bibitem[Kley et al.(2004)]{Kley04}
Kley, W., Peitz, J., \& Bryden, G.\ 2004, \aap, 414, 735

\bibitem[Marcy et al.(2002)]{Mar02}
Marcy,G. W., et al. 2002, \apj, 581, 1375

\bibitem[McArthur et al. (2004)]{McA04}
McArthur, B.E., et al. 2004, \apj, 614, L81

\bibitem[Menou \& Tabachnik(2003)]{Men03}
Menou, K., \& Tabachnik, S.\ 2003, \apj, 583, 473

\bibitem[Murray & Dermott(1999)]{Mur99}
Murray, C. D., \& Dermott, S. F. 1999, Solar System Dynamics (New
York: Cambridge Univ. Press) (MD99)

\bibitem[Nagasawa \& Ida(2000)]{Nag00}
Nagasawa, M., \& Ida, S. 2000, AJ, 120, 3311

\bibitem[Nagasawa (2005)]{Nag05}
Nagasawa, M., Lin, D.N.C., \& Thommes, E. 2005, \apj, 635, 578

\bibitem[Pilat-Lohinger(2008)]{Pil08}
Pilat-Lohinger, E., et al. 2008, ApJ, 681, 1639

\bibitem[Raymond et al.(2008)]{Ray008}
Raymond, S.~N., Barnes, R., \& Gorelick, N.\ 2008, \apj, 689, 478

\bibitem[Rivera \& Haghighipour(2007)]{Riv07}
Rivera, E., \& Haghighipour, N.\ 2007, \mnras, 374, 599

\bibitem[Thommes et al.(2008)]{Thom2008}
Thommes, E.~W., et al. 2008, \apj, 675, 1538

\bibitem[Valenti et al. (2005)]{Val05}
Valenti, J.A., \& Fischer, D. A.  2005, \apjs, 159, 141

\bibitem[Voyatzis \& Hadjidemetriou(2006)]{Voy06}
Voyatzis, G., \& Hadjidemetriou, J.~D. 2006, CeMDA, 95, 259

\bibitem[Voyatzis(2008)]{Voy08}
Voyatzis, G.\ 2008, \apj, 675, 802

\bibitem[Wisdom & Holman(1991)]{Wis91}
Wisdom, J., \& Holman, M. 1991, \aj, 102, 1528

\bibitem[Zhou et al.(2004)]{Zhou04}
Zhou, L.-Y., et al. 2004, \mnras, 350, 1495

\end{thebibliography}
\end{document}